\documentclass[%
 reprint,
superscriptaddress,
nofootinbib,
 amsmath,amssymb,
 aps,
 prl,
floatfix,
]{revtex4-2}

\usepackage{graphicx}
\usepackage{dcolumn}
\usepackage{bm}

\usepackage{epsfig}
\usepackage{amsmath}
\input{epsf}
\usepackage{psfrag}
\usepackage{xcolor}
\usepackage{pdfpages}
\usepackage[normalem]{ulem}
\makeatletter
\AtBeginDocument{\let\LS@rot\@undefined}
\makeatother
\newcommand{\bea}{\begin{eqnarray}}
\newcommand{\eea}{\end{eqnarray}}

\newcommand{\nn}{\nonumber}

\begin{document}

\title{ Nucleon Energy Correlators for the Color Glass Condensate}

\author{Hao-Yu Liu}%
\affiliation{College of Mathematics and Physics, Beijing University of Chemical Technology, Beijing 100029, China}

\author{Xiaohui Liu}
 \email{xiliu@bnu.edu.cn}
 \affiliation{Center of Advanced Quantum Studies, Department of Physics, Beijing Normal University, Beijing, 100875, China}
 \affiliation{Key Laboratory of Multi-scale Spin Physics, Ministry of Education, Beijing Normal University, Beijing 100875, China}
 \affiliation{Center for High Energy Physics, Peking University, Beijing 100871, China}

\author{Ji-Chen Pan}%
\affiliation{Institute of High Energy Physics and School of Physics, Chinese Academy of
	Sciences, Beijing 100049, China}

\author{Feng Yuan}%
 \email{fyuan@lbl.gov}
\affiliation{Nuclear Science Division, Lawrence Berkeley National Laboratory, Berkeley, CA 94720, USA}%

\author{Hua Xing Zhu}%
 \email{zhuhx@zju.edu.cn}
\affiliation{Zhejiang Institute of Modern Physics, Department of Physics, Zhejiang University, Hangzhou, 310027, China}%

\begin{abstract}
We demonstrate the recently proposed nucleon energy-energy correlator (NEEC) $f_{\rm EEC}(x,\theta)$ can unveil the gluon saturation in the small-$x$ regime in $eA$ collisions. 
The novelty of this probe is that it is fully inclusive just like the deep-inelastic scattering (DIS), with no requirements of jets or hadrons but still provides an evident portal to the small-$x$ dynamics through the shape of the $\theta$-distribution. 
We find that the saturation prediction is significantly different from the expectation of the collinear factorization.  
\end{abstract}

\maketitle
 
\textbf{\emph{Introduction.}}
%
%
Small-$x$ gluon saturation~\cite{Gribov:1983ivg,Mueller:1985wy,Mueller:1989st,McLerran:1993ni,McLerran:1993ka,McLerran:1994vd} has been one of the central focuses in nuclear physics community in recent years and will be a major research area in the future Electron Ion Collider
(EIC)~\cite{Accardi:2012qut,AbdulKhalek:2021gbh,Proceedings:2020eah}. An effective field theory called color-glass-condensate (CGC)~\cite{McLerran:1993ni,McLerran:1993ka,McLerran:1994vd} has been established to compute  
the hadronic and nuclear structure functions in deep inelastic scattering (DIS) at small values of Bjorken-$x_B$~\cite{Gelis:2010nm,Iancu:2003xm}. The CGC predicts the gluon saturation with a characteristic scale $Q_s$, as a consequence of the small-$x$ nonlinear dynamics governed by the BK-JIMWLK equation~\cite{Balitsky:1995ub,Kovchegov:1999yj,Jalilian-Marian:1997qno,Jalilian-Marian:1997jhx,Iancu:2000hn,Ferreiro:2001qy}. The saturation scale $Q_s$ represents the typical size of the gluon transverse momentum inside the nucleus and grows as the momentum fraction $x\to 0$. For large nucleus and small-$x$, typically $Q_s > \Lambda_{\rm QCD}$.  


Previous experiments from DIS in $ep$ collisions at HERA and hadron productions in $pA$ collisions at RHIC and LHC have shown some evidence of gluon saturation at small-$x$. With the planned EIC in the horizon, this physics will be explored in a systematic manner with unprecedented precision~\cite{Accardi:2012qut,AbdulKhalek:2021gbh,Proceedings:2020eah}. Extensive studies have been carried out for the EIC experiments, including the inclusive DIS structure functions at small-$x_B$~\cite{Albacete:2010sy,Lappi:2013zma,Ducloue:2017ftk} and the azimuthal correlations of di-jet/di-hadron/photon-jet/lepton-jet in the inclusive or diffractive processes~\cite{Dominguez:2010xd,Dominguez:2011wm,Mueller:2013wwa,Metz:2011wb,Dominguez:2011br,Dumitru:2015gaa,Dumitru:2016jku,Boer:2016fqd,Beuf:2017bpd,Dumitru:2018kuw,Mantysaari:2019hkq,Zhao:2021kae,Boussarie:2021ybe,Caucal:2021ent,Zhang:2021tcc,Taels:2022tza,Caucal:2022ulg,Boussarie:2014lxa,Boussarie:2016ogo,Salazar:2019ncp,Boussarie:2019ero,Boer:2021upt,Iancu:2021rup,Iancu:2022lcw,Hatta:2016dxp,Altinoluk:2015dpi,Mantysaari:2019csc,Hagiwara:2021xkf,Zheng:2014vka,Bergabo:2021woe,Bergabo:2022tcu,Iancu:2022gpw,Fucilla:2022wcg,Kolbe:2020tlq,Tong:2022zwp,Altinoluk:2022jkk}. These processes are considered as promising channels to look for the gluon saturation in $eA$ collisions.

In this manuscript, we present a novel approach to probe the gluon saturation in $eA$ collisions in terms of
the nucleon energy-energy correlator (NEEC) recently proposed in Ref.~\cite{Liu:2022wop}, 
which is an extension of the EEC~\cite{Basham:1978bw,Basham:1978zq} to the nucleon case. The EEC is the vacuum expectation of a set of final state correlators 
to reformulate jet substructures~\cite{Chen:2020vvp,Hofman:2008ar,Belitsky:2013ofa,Belitsky:2013xxa,Kologlu:2019mfz,Korchemsky:2019nzm,Dixon:2019uzg,Chen:2019bpb,Chen:2020adz,Chang:2020qpj,Li:2021zcf,Jaarsma:2022kdd,Komiske:2022enw,Holguin:2022epo,Yan:2022cye,Chen:2022jhb,Chang:2022ryc,Chen:2022swd,Lee:2022ige,Larkoski:2022qlf,Ricci:2022htc,Yang:2022tgm,Andres:2022ovj,Chen:2022pdu,Craft:2022kdo}, while the NEEC is the nucleon expectation of the initial-final state correlator. The latter encodes the partonic angular distribution induced by the intrinsic transverse momentum within the nucleon~\cite{Liu:2022wop}. Therefore we expect the features of the gluon saturation, especially the saturation scale $Q_s$ that measures the size of the intrinsic transverse momentum, should be naturally imprinted in the NEEC. Our numeric results in Figs.~\ref{fg:sig} and~\ref{fg:RAp} will show that the saturation predictions have distinguished behaviors as compared to those from the collinear factorization. From this comparison, we can further deduce the saturation scales in $ep$ and $eA$ collisions, respectively.
%

The quark contribution to the NEEC in the momentum space is defined as 
\bea\label{eq:feec} 
 f_{q,\rm EEC}(x,\theta)
&=& 
\int \frac{dy^-}{4\pi E_A} 
e^{-i x P\, y^-} \nn \\ 
&& \hspace{-5.ex} 
\times  
\gamma^+
\langle A|
{\bar \psi}(y^-) {\cal L}^\dagger (y^-) \, 
{\cal E}(\theta) {\cal L}(0)
\psi(0)
| A \rangle \,,     
\eea 
%
where $x$ is the momentum fraction that initiates a scattering process, meanwhile we measure the energy deposit in a detector at a given angle $\theta$ from the initial state radiation and the remnants through the energy operator, ${\cal E}(\theta) |X\rangle = 
\sum_{i\in X} E_i \delta(\theta_i^2 - \theta^2 )  | X\rangle $~\cite{Sveshnikov:1995vi,Tkachov:1995kk,Korchemsky:1999kt,Bauer:2008dt}\,.  
The measured energy deposit is normalized to the energy $E_A$ carried by the nucleus $A$. Here, $\psi$ is the standard quark fields and 
${\cal L}$ is the gauge link. The gluon EEC can be defined similarly.  When $\theta E_A \sim \Lambda_{\rm QCD}$, the $f_{\rm EEC }$ probes the intrinsic transverse
dynamics of the nucleus $A$ through the operator ${\cal E}(\theta)$. 

In the collinear factorization, when $\theta E_A \gg \Lambda_{\rm QCD}$, the $f_{q,\rm EEC}(x,\theta)$ can be further factorized as
\bea\label{eq:feecfac} 
f_{i,\rm EEC}(x,\theta) 
= 
\int \frac{d\xi}{\xi}
I_{ij}\left(\frac{x}{\xi},\theta\right) \left[\xi f_{j/A}\left(\xi\right) \right]\,, 
\eea 
where $f_{j/A}(\xi)$ is the collinear PDF, and $I_{ij}$ is the matching coefficient. The detailed derivation of the factorization is given in~\cite{Cao:2023rga} and can be obtained by taking the derivative of the cumulant result in Eq.~(36) with respect to $\theta$  therein.  Here $I_{ij}$ is found to be solely determined by the vacuum collinear splitting functions~\cite{Liu:2022wop,Cao:2023rga}. 
\begin{figure}[htbp]
  \begin{center}
   \includegraphics[scale=0.195]{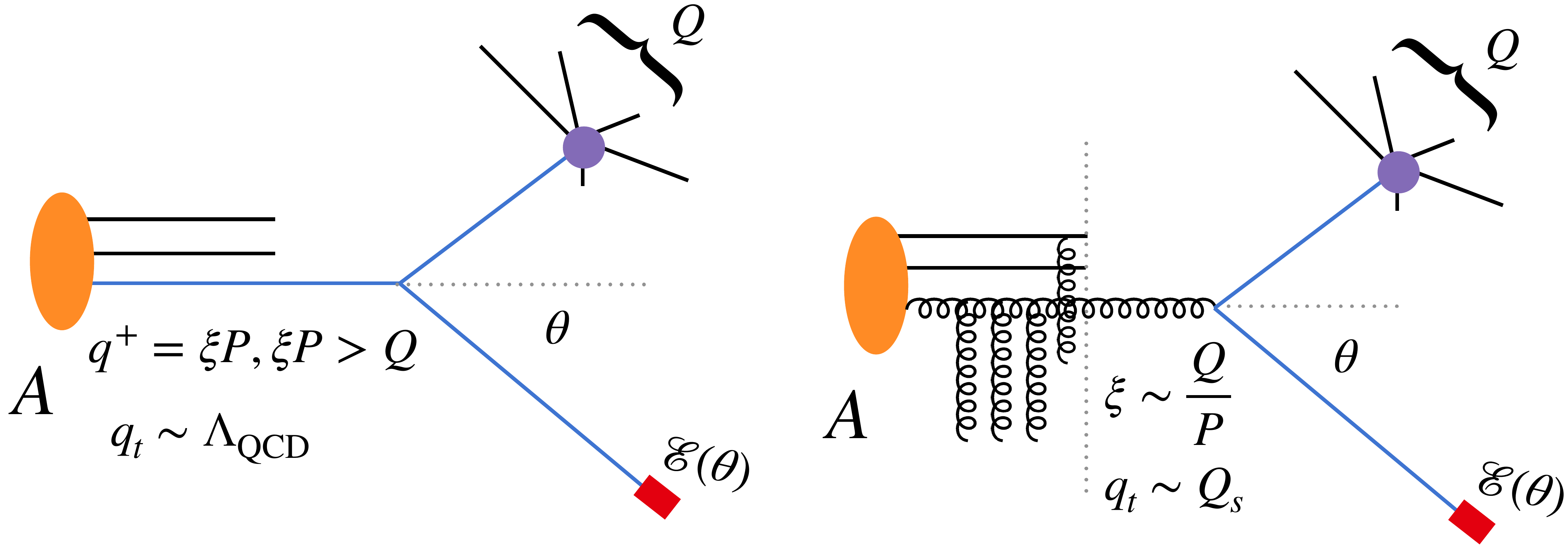} 
\caption{The $f_{EEC}(x,\theta)$ in the collinear factorization (left) and the CGC framework (right). Here $Q$ represents the center of mass energy of the partonic cross section.}
  \label{fg:collvscgc}
 \end{center}
\end{figure}

As the values of $x$ decrease, the $f_{q,\rm EEC}$ receives dramatically enhanced contributions from the low $x$ gluon. 
In this high gluon density regime, the appropriate factorization framework is the CGC formalism, where, especially, the multiple parton interaction effects have been systematically taken into account. 
As result, the shape of the $\theta$-distribution will be modified, due to a sizable initial transverse momentum $q_t$ of order the saturation scale $Q_s$, see, the illustrations in Fig.~\ref{fg:collvscgc}. Therefore, the NEEC can be used to probe the gluon saturation phenomenon and the small-$x$ dynamics, as we will show in the rest of this manuscript.

\begin{figure}[htbp]
  \begin{center}
   \includegraphics[scale=0.185]{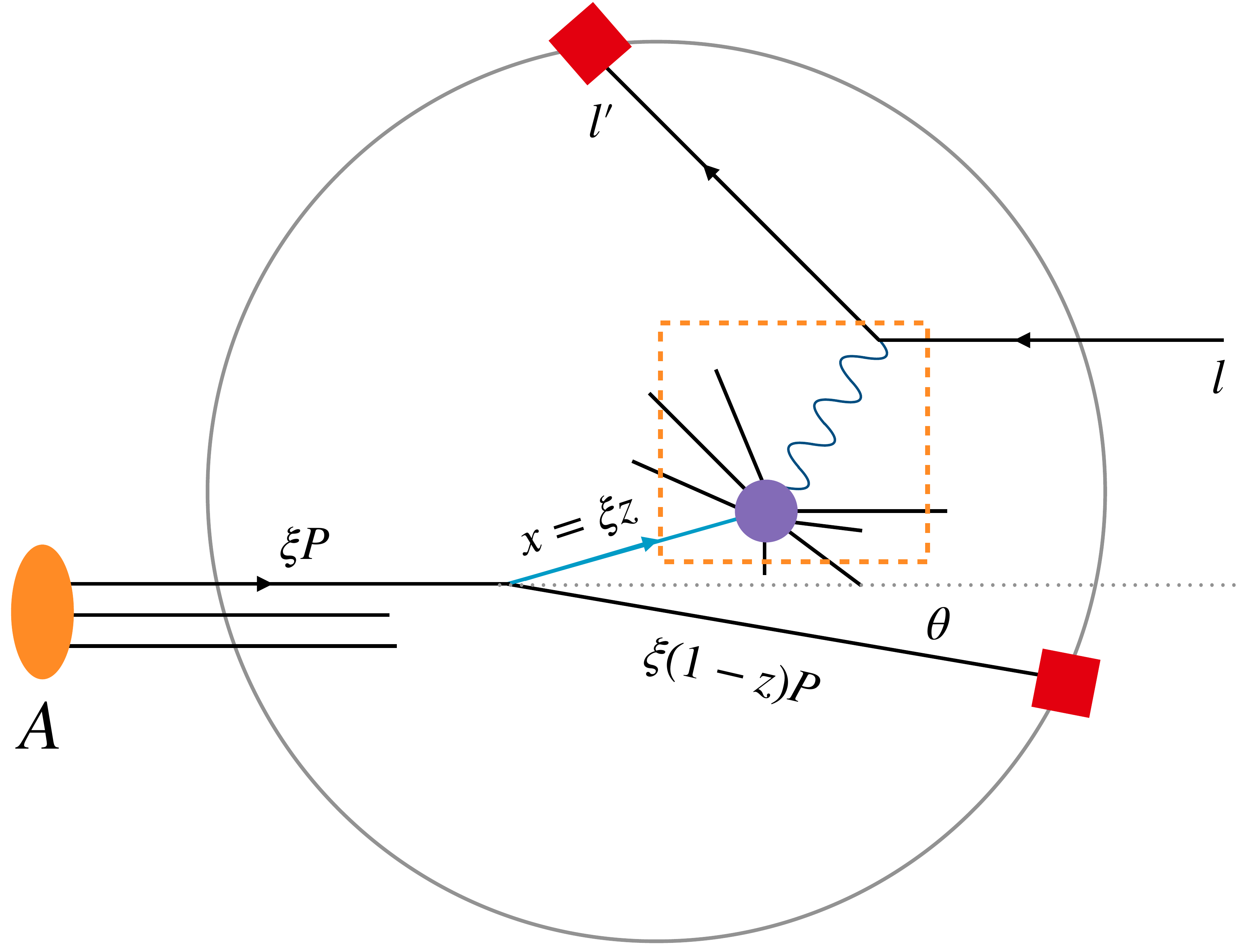} 
\caption{The $x_B$ and $Q^2$ measurement in DIS with a forward detector that records the energy flow $\sum_iE_i$ at the angle $\theta$. The leading contribution is also illustrated where the collinear splitting initiates the DIS process and a daughter parton that hits the detector at $\theta \ll 1$. The momentum fractions are shown. We abbreviate $P^+$ with $P$ in this work for simplicity notation. }
  \label{fg:measure}
 \end{center}
\end{figure}

\textbf{\emph{The measurement and the factorization theorem.}} We follow~\cite{Liu:2022wop} to consider the unpolarized DIS process $l + A \to l' + X $ in the Breit frame. We assume the nucleus is moving along the $+z$-direction. 
We measure the Bjorken $x_B = \frac{-q^2}{2P\cdot q }$, the photon virtuality $Q^2 = -q^2$ and the energy $\sum_i E_i$ that deposits in a calorimeter at an angle $\theta$ with respect to the beam, as shown in Fig.~\ref{fg:measure}. Here $q = l'-l$ is the momentum carried by the virtual photon. We then measure the weighted cross section $\Sigma(Q^2,x_B,\theta)$ defined as 
\bea\label{eq:eec-def1} 
\Sigma(Q^2, x_B, \theta)   
= \sum_i  \int  d\sigma(x_B,Q^2,p_i) \, 
 \frac{ E_i }{E_A}  \, \delta(\theta^2 - \theta_i^2  )
\,,
\eea
where $E_A$ is the energy carried by the incoming nucleus. We note that the energy weight suppresses the soft contributions, which is an important feature of the proposed measurement and its resulting NEEC. 

In order to probe the small-$x$ dynamics, we are particularly interested in the scenario in which $x_B \ll 0.1$, and we place the detector in the far-forward region such that $Q \theta \ll Q$ while $Q\theta \sim Q_s \gg \Lambda_{\rm QCD}$. At this point, we emphasize that the measurement involves neither additional hadron tagging nor jet clustering, and in contrast to the TMD which restricts the events in the small $q_t$ region, this approach is inclusive and does not veto events. It weights the full cross section by the energy recorded at a certain angle $\theta$, therefore the probe is as inclusive as the DIS but with additional control via $\theta$.

%
When $\theta Q \gg \Lambda_{\rm QCD}$, the weighted cross section can be calculated perturbatively in the collinear factorization. More interestingly, when $Q\theta \ll Q$, the $\Sigma(Q^2,x_B,\theta)$ fulfils the factorized form
\bea\label{eq:sigfac} 
\Sigma(Q^2,x_B, \theta) 
= \int \frac{dx}{x}
{\hat{\sigma}}_{i,{\rm DIS}}\left(\frac{x_B}{x},Q\right) f_{i,{\rm EEC}}(x,\theta) \,, 
\eea 
where 
$\hat{\sigma}_{i,{\rm DIS}}$ is the fully inclusive partonic DIS cross section.
$f_{i,{\rm EEC}}$ is the NEEC in Eq.~(\ref{eq:feec}). 
The factorization theorem can be seen from Ref.~\cite{Cao:2023rga} by taking out the $x_B^{N-1}$ weight.
The $\theta$-dependence enters entirely through the NEEC $f_{\rm EEC}(x,\theta)$, and therefore the $\theta$ distribution of the $\Sigma(Q^2,x_B,\theta)$ probes the NEEC when $\theta$ is small. We note that $f_{\rm EEC}$ satisfies the same collinear evolution as the collinear PDFs
\bea\label{eq:ll-resm} 
\frac{df_{i,{\rm EEC}}(x,\theta)}{d\ln\mu} 
= P_{ij} \otimes f_{j,{\rm EEC}} \,, 
\eea  
as required by $d\Sigma/d\ln \mu = 0$, and since $d{\hat \sigma}_{i,{\rm DIS}}/d\ln \mu = - P_{ji}\otimes \hat{\sigma}_{j,{\rm DIS}}$. Here the convolution in the momentum fraction is defined as
$f \otimes g(x) \equiv  
\int_x^1 \frac{dz}{z} f\left(\frac{x}{z}\right)g(z)
$. It is clear from the evolution that there is no perturbative Sudakov suppression in $f_{\rm EEC}$, due to the absence of the soft contribution in the collinear factorization eliminated by the energy weight~\cite{Liu:2022wop}. 
%
%

Here are some comments on the factorization:     


A key feature of the NEEC is that it does not involve TMDs and in the kinematic region of $Q\theta \gg \Lambda_{\rm QCD}$, the $\theta$ distribution comes entirely from collinear splitting, as illustrated in Fig.~\ref{fg:measure}. As demonstrated in~\cite{Liu:2022wop}, soft gluon radiations do not contribute to the NEEC and there are no Sudakov double logs from perturbative calculations which are the key elements to apply the TMD factorization properly. One way to see this is that the energy weight in Eq.~(\ref{eq:eec-def1}) suppresses the soft contributions.
Another way is to notice that the measurement under consideration does not restrict the transverse momentum of parton (in the blue line in Fig.~\ref{fg:measure})  initiating the hard interaction to be small, which is different from the SIDIS but a lot like the inclusive DIS (highlighted by the dashed box in Fig.~\ref{fg:measure}). 
The factorized form in Eq.~(\ref{eq:sigfac}) also manifests the similarity between the NEEC observable and the inclusive DIS structure functions.

The leading order contribution to the measurement in Fig.~\ref{fg:measure} can be found in the Supplemental Material of Ref.~\cite{Liu:2022wop} where one considers a parton out of the nucleus $A$ with momentum $\xi P$ splits into a parton with momentum fraction $(1-z)\xi$ that hits the detector at $\theta$. 
At LO $z = \frac{x_B}{\xi}$ is the momentum fraction with respect to the incoming parton and $k_t = \frac{1}{2} \xi(1-z)P \, \theta $ is the transverse momentum of the detected parton. 
Note that $k_t$ is insensitive to the initial parton transverse momentum, for $\xi P\theta \gg \Lambda_{\rm QCD}$. 
When $x_B \ll 0.1$, the gluon density is overwhelmingly dominant and we found for $\Lambda_{\rm QCD} \ll \theta Q \ll Q$ that 
%
 $\Sigma(Q^2,x_B,\theta) = \sum_q 
 \frac{4\pi\alpha^2 e_q^2}{Q^4} \, 
 f_{q,{\rm EEC}}(x_B,\theta)$,
with
\bea\label{eq:small-x-coll}
&& f_{q,{\rm EEC}}(x,\theta) \nn \\ 
 &=&  \frac{\alpha_sT_R}{2\pi \theta^2}  \int_x^1 \frac{d\xi}{\xi}
 (1-\xi) (\xi^2+(1-\xi)^2)  
\, \left[\frac{x}{\xi} f_g\left(\frac{x}{\xi}\right)\right]
 \,.
\eea

The collinear factorization predicts a $\frac{1}{\theta^2}$-scaling behavior
at ${\cal O}(\alpha_s)$. For very small $\theta$, the scaling rule could receive corrections from both the evolution of the $f_{\rm EEC}$ in Eq.~(\ref{eq:ll-resm}) and non-perturbative effects. But for generic small $\theta$, these effects are mild and therefore $\theta^2\Sigma$ will be insensitive to the values of $\theta$, up to ${\cal O}(\theta Q)$ power corrections. Furthermore, since the energy weight kills the soft contribution, to all orders there will be no perturbative Sudakov suppression in the small $\theta$ region in the collinear factorization~\cite{Liu:2022wop}, as is clear from Eq.~(\ref{eq:ll-resm}).

In practice, we demand $\theta > 0.1$, corresponding to a rapidity range of $|y|<3$. This ensures that the detected forward partons are well-separated from the target beam. We thus expect high twist effects in the collinear factorization, such as interactions between the beam spectator and the detected parton, to be mild. Consequently, their corrections to the predicted $\theta$ distribution are small.  
However, if there exists a large saturation scale $Q_s$ as predicted by the CGC for small $x_B$, the $\theta$ behavior could change dramatically, as we will demonstrate.



%
\begin{figure}[htbp]
  \begin{center}
   \includegraphics[scale=0.5]{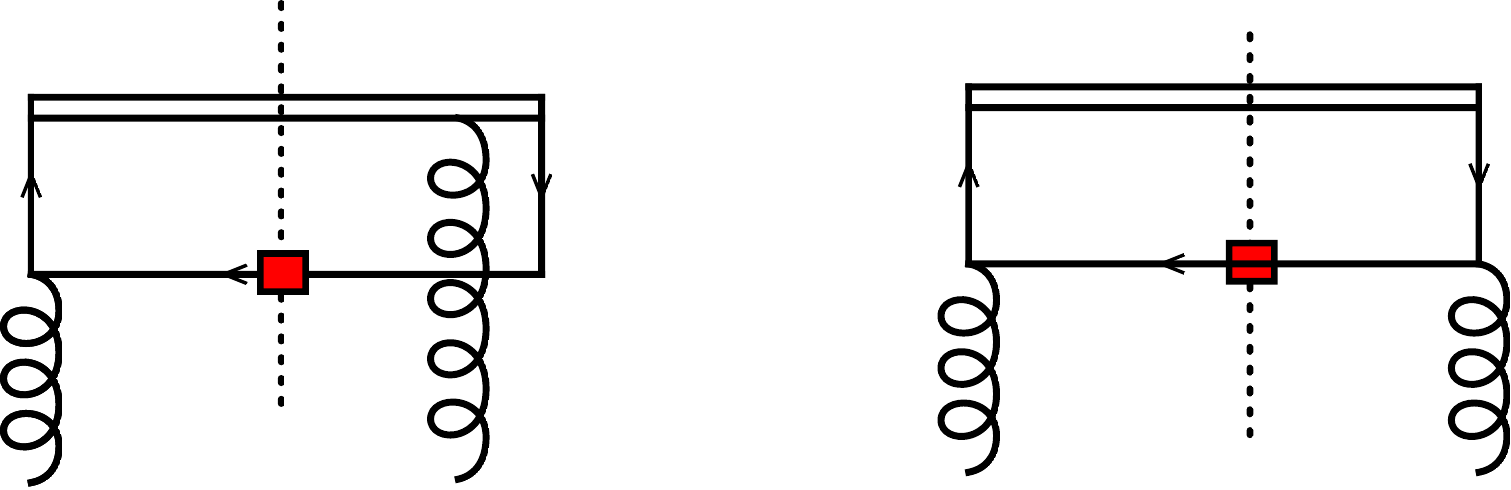} 
\caption{Representative diagrams for the leading contribution to $f_{q,{\rm EEC} }(x,\theta)$ in the small-$x$ region, where the double line represents the gauge link and the gluon requires momentum $g^+=x_g P$ and $g_t \sim Q_s \sim \theta Q$.}
  \label{fg:eec-small-x}
 \end{center}
\end{figure}

\textbf{\emph{The NEEC in the small-$x$ regime.}}
In the small-$x$ region, the gluon density grows as $\frac{1}{x}$ and becomes overwhelmingly important and has to be resummed to all orders. To realize such resummation in $f_{\rm EEC}$, we invoke the CGC effective theory framework and  follow the strategy in~\cite{Marquet:2009ca, Xiao:2017yya,Zhou:2018lfq} to write the NEEC in terms of the CGC dipole distribution~\footnote{The complete calculation using the full dipole amplitude $\psi_{T,L}^{\gamma^\ast\to q{\bar q}}$ for $\gamma^\ast \to q{\bar q}$ is presented in the Supplemental Material. Both approaches agree in the small $\theta$ limit.}. By evaluating the diagrams in Fig.~\ref{fg:eec-small-x}, we find in the leading logarithmic (LL) approximation
\bea \label{eq:feecx}
f_{q,{\rm EEC}}(x_B,\theta)
&=& \frac{N_C S_\perp}{8\pi^4} 
 \int d^2 \vec{g}_t \nn \\ 
&& \hspace{-8.ex} \times
\int_{\xi_{\rm cut}}^{1}  
\frac{d\xi }{\xi}
{\cal A}_{qg}\left(\xi,\theta,\vec{g}_t\right)
\,  F_{g,x_B}(\vec{g_t}) \,, 
\eea 
where $S_\perp$ is the averaged transverse area of the target nucleus and $g_t \sim Q_s \sim \theta Q$ is the transverse momentum transfer. $F_{g,x_F}  = \int\frac{d^2\vec{r}}{4\pi^2 }
e^{-i\vec{g}_t\cdot \vec{r}_t}S_{x_F}^{(2)}(\vec{r}_t)  $ is the CGC dipole distribution  evaluated at the scale $x_F$,  
where 
$
S^{(2)}_{x_F}(\vec{r}_t)  = 
\frac{1}{N_C} \langle  {\rm Tr}[W(\vec{r}_t)W^\dagger(\vec{0})]\rangle_{x_F}$, with the Wilson line $W(\vec{r}_t)$ constructed out of the gauge field $A_c$, $W(\vec{r}_t)=\mathcal{P}\exp\left\{ig\int_{-\infty}^{+\infty}dx^+T^cA_c^-(x^+,\vec{r}_t)\right\}$. $x_F \frac{Q}{x_B}$ is the rapidity scale/boundary that separates the fast-moving modes being integrated out and the active slow-moving partons in the CGC effective framework. In this work, we default to the natural choice $x_F = x_B$. $\frac{1-\xi}{\xi} Q$ is the momentum ``$+$"-component that enters the detector. $\xi_{\rm cut}$ is determined by requiring the momentum of the active quark does not exceed the rapidity boundary. 
 Here the coefficient ${\cal A}_{qg}$ is given by
\bea\label{eq:Asmall-x} 
{\cal A}_{qg}(\xi, \theta,\vec{g}_t) 
&=& \frac{1}{ \theta^2}
(1-\xi) 
 \vec{k}_t^2  
(\vec{k}_t - \vec{g}_t)^2
\nn \\   
&& \hspace{-10.ex} \times 
\, 
\left| \frac{\vec{k}_t}{\xi \vec{k}_t^2+(1-\xi) (\vec{k}_t-\vec{g}_t)^2 } - \frac{\vec{k}_t-\vec{g}_t}{(\vec{k}_t-\vec{g}_t)^2} \right|^2 \,, \quad  
\eea 
with $k_t$ defined as 
$k_t = \frac{1-\xi}{\xi}\frac{Q}{2} 
\theta$, should be of order $Q_s$. 

It is easy to show that if $g_t\sim Q_s \ll Q\theta $, Eq.~(\ref{eq:feecx}) 
reduces to the $\frac{1}{\theta^2}$-scaling behavior of the collinear factorization in Eq.~(\ref{eq:small-x-coll}). On the other hand, if $\theta Q \ll Q_s$, 
Eq.~(\ref{eq:feecx}) scales as $\theta^0$. 
We thus expect that in CGC, the $\theta^2\Sigma$ will be independent of the $\theta$ for $\theta Q \gg Q_s $, however,  contrary to the collinear factorization, suppressed when $\theta Q \ll Q_s$. Meanwhile, the $\theta$ region between these two limits provides the opportunity to estimate the saturation scale $Q_s$. One thing to mention is that in practice we will focus on $\theta > 0.1$, therefore although the value is small, it is still sufficiently separated from the beam. Given that the beam remnants are typically very energetic in the small $x_B$ region, it is unlikely that they will hit the detector at such angles.

\textbf{\emph{Numerics.}} Now we study the numerical impacts of the small-$x$ dynamics on the shape of the $\theta^2\Sigma(Q^2,x_B,\theta)$ distribution from Eq.~(\ref{eq:feecx}), compared with the collinear prediction. We are particularly interested in the region $\theta \ll 1$ where the $\theta$ distribution probes direcly the $f_{\rm EEC}(x,\theta)$, see Eq.~(\ref{eq:sigfac}). 
For the small-$x$ dipole distribution $S_{x_F}^{(2)}(\vec{r}_t)$, we use both the MV model with rcBK running~\cite{Balitsky:1995ub,Kovchegov:2006wf,Kovchegov:1999yj,Kovchegov:2006vj,Albacete:2010sy,Golec-Biernat:2001dqn,Albacete:2007yr,Balitsky:2006wa,Gardi:2006rp,Balitsky:2007feb,Berger:2010sh} and the GBW model~\cite{Golec-Biernat:1998zce}. 

As for the MV model with rcBK running, we adopt the MV-like model~\cite{Fujii:2013gxa} as the initial condition, whose form  is
$S^{(2)}_{x_0}(\vec{r}_t) = \exp \left[-\frac{(r_t^2 Q_{s0}^2)^\gamma}{4}\ln\left(\frac{1}{\Lambda r_t}+e\right)\right]
$, 
where we choose $x_0=0.01$, $\gamma=1.119,\Lambda=0.241~\text{GeV},Q_{s0}^2=A^{1/3} 0.168~\text{GeV}^2$ with $A$ the atomic number. 
 We use the solution to the LL BK evolution with $\alpha_s$ running~\cite{Kovchegov:2006vj,Balitsky:2006wa,Fujii:2013gxa} of the dipole distribution to evolve the dipole distribution from $x_0$ to $x_F$.  In our calculation, we use the result fitted from the HERA data for the transverse area of the nucleus $S_\perp$~\cite{Lappi:2013zma}. The GBW model is implemented using
$S^{(2)}_{x_F}(\vec{r}_t) =  \exp \left[-\frac{1}{4}r_t^2 Q_s^2(x_F)\right]$,  
where $Q_s^2(x_F) = A_N (x_0/x_F)^{\lambda}\>{\rm GeV}^2$ and we use $x_0=2.24\times 10^{-4}$, $\lambda=0.27$ and $A_N=1$ for the proton while $A_N=5$ for the Au~\cite{Golec-Biernat:2017lfv}. 

\onecolumngrid
\begin{widetext}
\begin{figure}[htbp]
  \begin{center}   
    \includegraphics[scale=0.47]{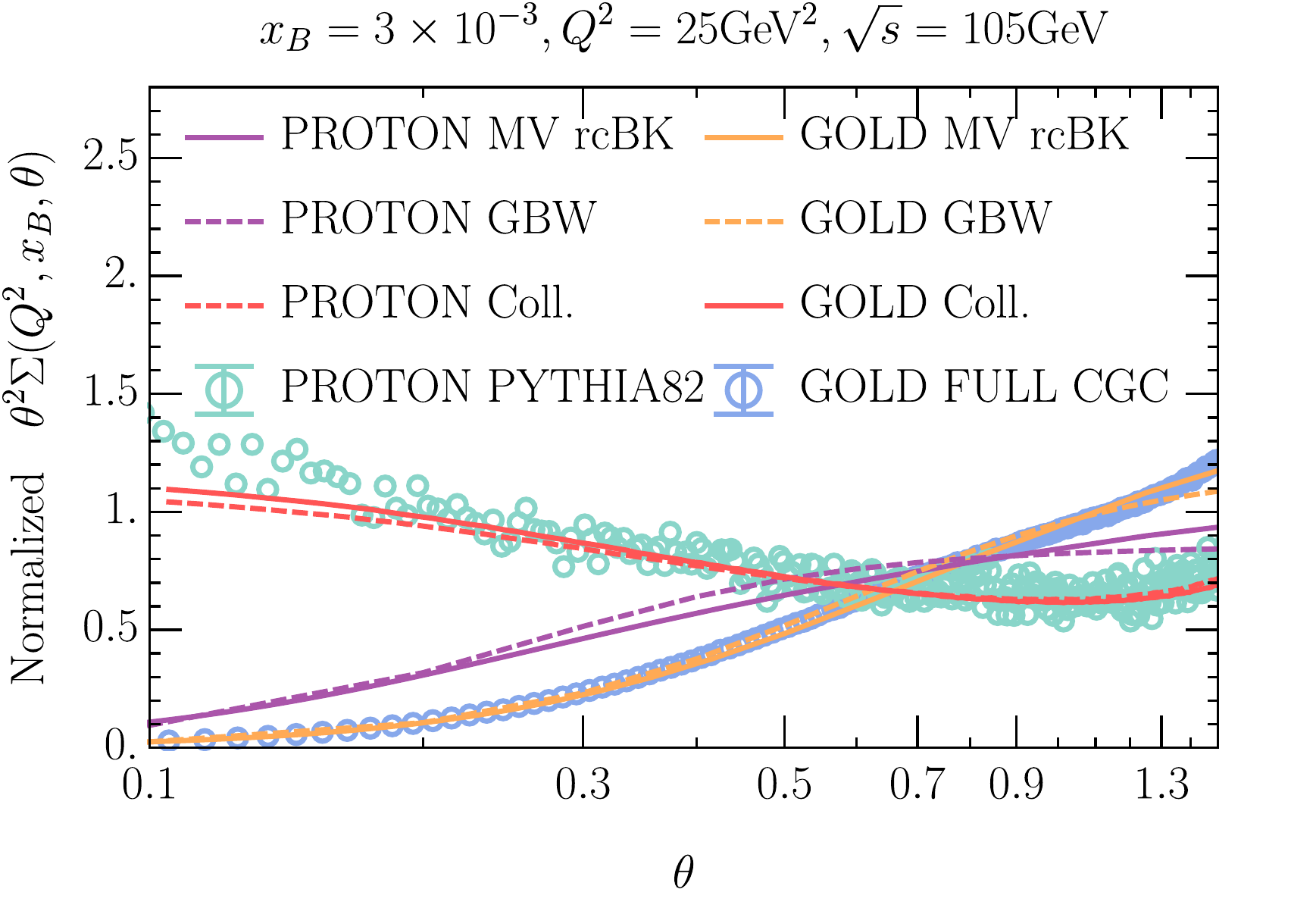} 
      \hspace{\fill}
    \includegraphics[scale=0.47]{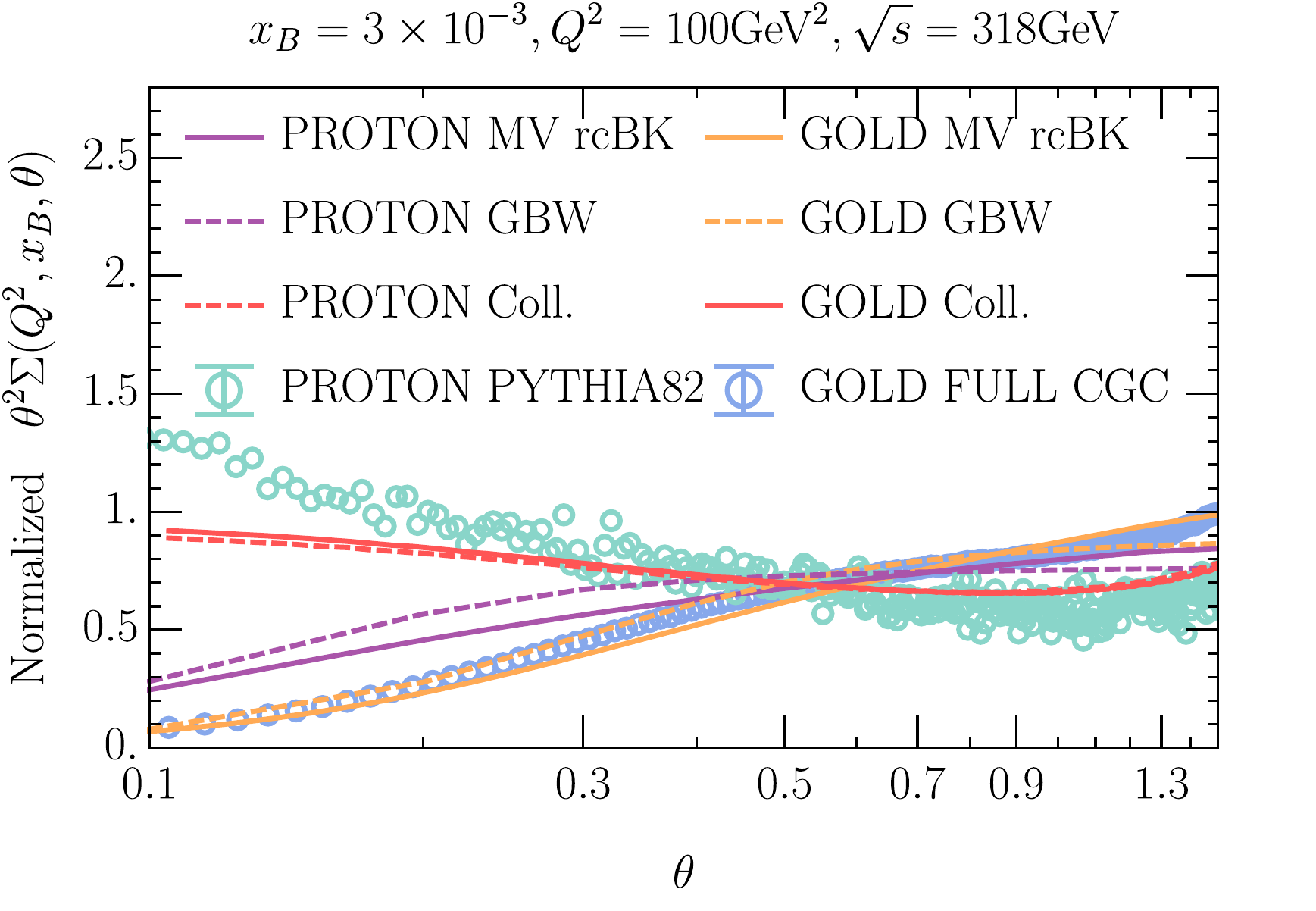}
\caption{
Normalized $\theta^2\Sigma(Q^2,x_B,\theta)$ distribution for proton and Au, with $x_B = 0.003$, for both $Q^2 = 25\,{\rm GeV}^2\,, \sqrt{s} = 105\, {\rm GeV}$ (left panel) and $Q^2 = 100\,{\rm GeV}^2\,, \sqrt{s} = 318\, {\rm GeV}$ (right panel). Green dots are from {\tt Pythia82} simulation, in which we demand $0.00295< x_B < 0.00305$, and $25 {\rm GeV}^2\le Q^2<35 {\rm GeV}^2 $ for the left panel while $100{\rm GeV}^2\le Q^2<110{\rm GeV}^2$ for the right.}
  \label{fg:sig}
 \end{center}
   \vspace{-5.ex}
\end{figure}
\end{widetext}
\twocolumngrid
In Fig.~\ref{fg:sig}, we show the CGC predictions for $\theta^2 \Sigma(Q^2,x_B,\theta)$ as a function of $\theta$. Since we are only interested in the shape, we normalized the distribution by $\int_{\theta_{\rm min}}^{\theta_{\rm max}} d\theta \theta^2\Sigma$. We fixed $x_B = 3\times 10^{-3}$ and choose $Q^2 = 25\, {\rm GeV}^2$, $\sqrt{s} = 105\, \rm{GeV}$ (left panel) and $Q^2 = 100\, {\rm GeV}^2$, $\sqrt{s} = 318\, \rm{GeV}$ (right panel). We present predictions from CGC for both proton (in purple lines) and Au (in orange), by the MV model with rcBK running and GBW model. 
We see both models predict similar shapes in the $\theta$ spectrum, in which the small-$\theta$ region is suppressed. They are impressively different from the collinear expectations (in red lines and green dots). In the figure, the collinear predictions (red lines) are made out of the complete fixed order $\alpha_s$ calculation, without the  $Q\theta \ll Q$ approximation, using {\tt CT18A}~\cite{Hou:2019efy} and {\tt EPPS21}~\cite{Eskola:2021nhw} PDF sets for proton and Au, respectively. To validate our collinear calculation and to estimate the size of the evolution effect in Eq.~(\ref{eq:ll-resm}), we also run a {\tt Pythia82} simulation~\cite{Sjostrand:2014zea} for the proton case, where the LL resummation is performed. We see that for large $\theta$, the fixed order calculations agree well with the the {\tt Pythia} simulation, while for small $\theta$ values, the resummation effects could be sizable but does not suppress the small-$\theta$ region due to the absence of the perturbative Sudakov factor in $f_{\rm EEC}$ in the collinear factorization.  
The collinear prediction for the Au follows closely the proton's. 
The notable difference demonstrates that the $f_{\rm EEC}(x,\theta)$ could serve as a clean probe of the small-$x$ phenomenon. For comparison, we also show the predictions from the full CGC calculation derived in the Supplemental Material~\cite{supp} using the GBW model in purple circles. We also find that the $\theta$ shapes predicted by the rcBK evolution are similar to the predictions by the linearized evolution, rcBFKL~\cite{Kovchegov:2006wf} with the {\it same gluon saturation initial condition}. This can be understood that for this kinematics the evolution effect is mild and we are probing the onset of the gluon saturation.

In Fig.~\ref{fg:sig}, the proton spectrum turns into a plateau for large values of $\theta$, which is expected from Eq.~(\ref{eq:feecx}) when $Q\theta \gg Q_s$. We can define a turning point around which the slope of the distribution starts to switch its monotonicity. The turning point allows us to estimate the size of the saturation scale $Q_s$. For instance, from the left panel of Fig.~\ref{fg:sig}, the turning point for the proton is roughly around $\theta \sim 0.15 - 0.2$ and thus $Q_s \sim \theta Q \sim 0.75 - 1.0 \, {\rm GeV}$ which is consistent with the values of the proton $Q_s$. And we estimate the saturation scale for the Au will be around $\theta\sim 0.4-0.5$ and thus $Q_s \sim \theta Q \sim 2-2.5 \, {\rm GeV}$. The right panel of Fig.~\ref{fg:sig} is similar to the left, but with $Q^2 = 100\,{\rm GeV}^2$. Since $Q^2$ is larger, the distribution enters the plateau earlier as expected.  Now the turning point for the Au is around $\theta \sim 0.2 - 0.3$ which again indicates that $Q_s \sim 2 - 3\, {\rm GeV}$, consistent with the $Q^2 = 25\, {\rm GeV}^2$ case. 

We can further introduce the nuclear modification factor  $R_{pA} = \frac{ A^{-1} \Sigma_A(Q^2,x_B,\theta) }{ \Sigma_p(Q^2,x_B,\theta)}$, which helps to reduce the systematics. In the collinear factorization, for $\theta Q \gg \Lambda_{\rm QCD}$, the $\theta$ distribution is determined by the matching coefficient $I_{ij}$ as predicted by Eq.~(\ref{eq:small-x-coll}), which is independent of the incoming nucleus species. Thus taking the ratio $R_{Ap}$ reduces the impacts from perturbative higher order corrections as well as possible non-perturbative hadronization effects, and the collinear factorization predicts the $R_{Ap}$ insensitive to the $\theta$ values, as showed explicitly as red lines in Fig.~\ref{fg:RAp}.    
 %
%
\begin{figure}[htbp]
  \begin{center}   \includegraphics[scale=0.45]{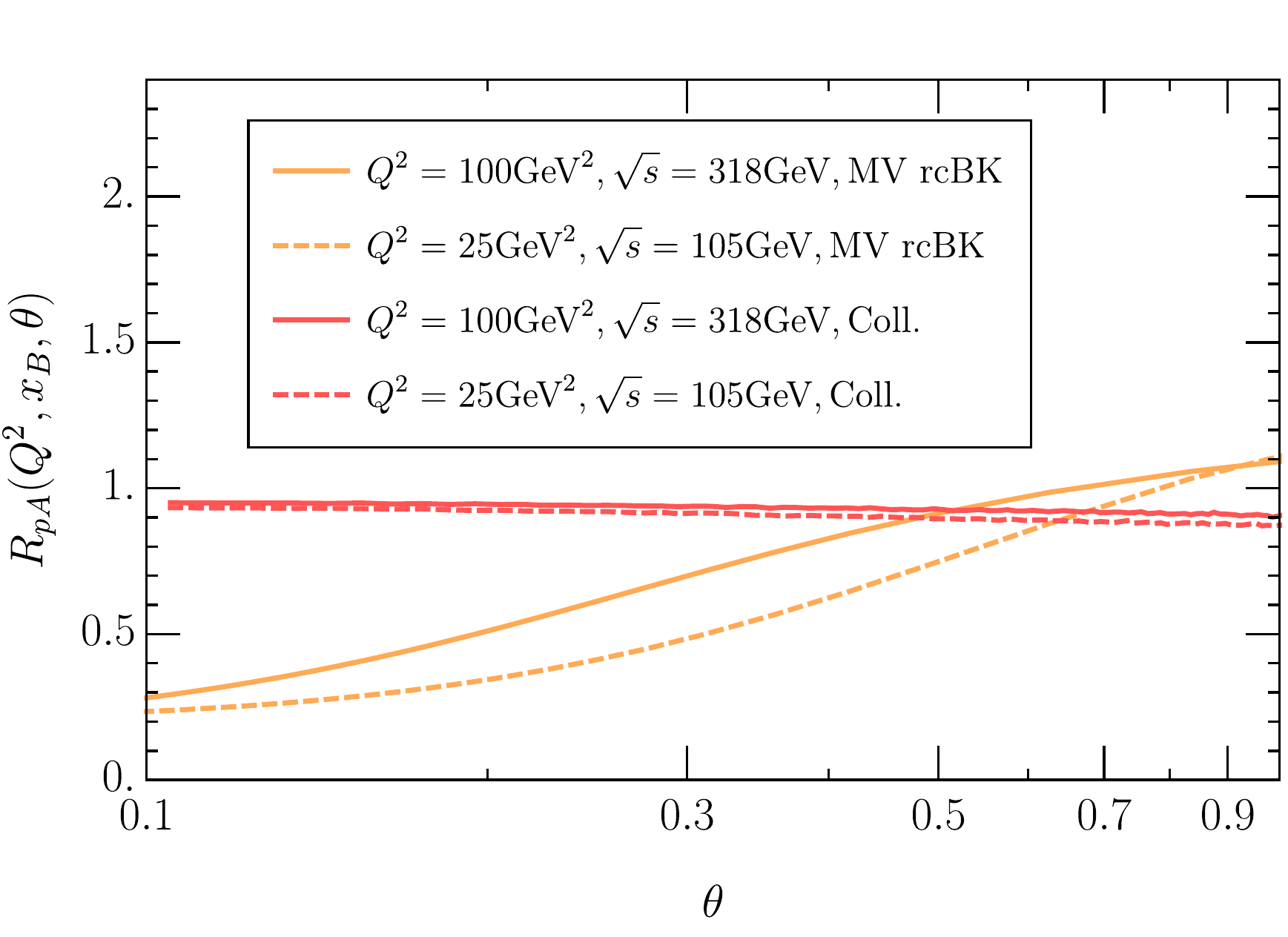} 
\caption{$R_{pA}$ as a function of $\theta$, with $x_B=3\times 10^{-3}$ using the MV model with rcBK running and collinear factorization.}.
  \label{fg:RAp}
 \end{center}
   \vspace{-5.ex}
\end{figure}

Once again, the small-$x$ formalism changes the pattern as we observed in Fig.~\ref{fg:RAp}, where the modification factor $R_{pA}$ is suppressed in the small $\theta$ region, while converges toward around unity as $\theta$ becomes large and $Q \theta \gg Q_s$.



\textbf{\emph{Conclusions.}} 
In this manuscript, we have proposed the nucleon energy-energy correlator (NEEC) as a new probe of the gluon saturation phenomenon in DIS at the 
future electron-ion colliders. In
particular, we have shown that the $\theta$-shape of the NEEC $f_{\rm EEC}(x,\theta)$ behaves differently in the collinear factorization theorem and the CGC formalism. The drastic difference is due to the intrinsic transverse momentum of order $Q_s$ induced by the non-linear small-$x$ dynamics. We thus expect the $f_{\rm EEC}$ to complement the other standard small-$x$ processes and offer a great opportunity to pin down the onset of the gluon saturation phenomenon in $ eA$ collisions. 

The NEEC probe has an advantage over other standard small-$x$ processes because it is fully inclusive and does not involve fragmentation functions or jet clustering. This makes the observable both theoretically and experimentally clean. Further extensions to other observables induced by the intrinsic transverse dynamics of the nucleon/nucleus are expected. Similar measurements, such as measuring NEEC in prompt photon production, can also be carried out at the LHC and fit into the ALICE forward calorimeter program~\cite{ALICE:2020mso}. We hope that our results will motivate the proposed measurement at current and future electron-ion facilities and stimulate further applications of NEEC in nuclear structure studies.



\begin{acknowledgments}
 \textbf{\emph{Acknowledgement.}} 
 We are grateful to Farid Salazar, Hongxi Xing, Jian Zhou for useful discussions.
This work is supported by the Natural Science Foundation of China under contract No.~12175016 (X.~L.), No.~11975200 (H.~X.~Z.) and the Office of Science of the U.S. Department of Energy under Contract No. DE-AC02-05CH11231 (F.Y.). The work of J.~C.~P. is supported in part by the National Natural Science Foundation of China under Grants No. 11925506, and No. 11621131001 (CRC110 by DFG and NSFC).
\end{acknowledgments}




\bibliographystyle{h-physrev}   
\bibliography{refs}

\clearpage


\section*{Supplementary material}

# Supplemental Materials for "Nucleon Energy Correlators for the Color Glass Condensate"

Hao-Yu Liu,[1] Xiaohui Liu,[2,3,4,*] Ji-Chen Pan,[5,6] Feng Yuan,[7,†] and Hua Xing Zhu[8,‡]

[1] College of Mathematics and Physics, Beijing University of Chemical Technology, Beijing 100029, China
[2] Center of Advanced Quantum Studies, Department of Physics,
Beijing Normal University, Beijing, 100875, China
[3] Key Laboratory of Multi-scale Spin Physics, Ministry of Education,
Beijing Normal University, Beijing 100875, China
[4] Center for High Energy Physics, Peking University, Beijing 100871, China
[5] Institute of High Energy Physics, Chinese Academy of Sciences, Beijing 100049, China
[6] School of Physics, University of Chinese Academy of Sciences, Beijing 100049, China
[7] Nuclear Science Division, Lawrence Berkeley National Laboratory, Berkeley, CA 94720, USA
[8] Zhejiang Institute of Modern Physics, Department of Physics, Zhejiang University, Hangzhou, 310027, China
(Dated: May 12, 2023)

Here we present the $\Sigma(Q^2, x_B, \theta)$

$$\Sigma(Q^2, x_B, \theta) = \sum_i \int d\sigma(x_B, Q^2, p_i) \frac{E_i}{E_A} \delta(\theta^2 - \theta_i^2), \tag{1}$$

within the CGC formalism without the small $\theta$ approximation. The calculation is achieved by evaluating amplitude for a virtual photon $\gamma^*$ splitting into the $q\bar{q}$ pair, as shown in Fig. 1. Each of the quarks will enter the detector to contribute to the energy deposit $E_i$, which we assume it is always the $\bar{q}$. The contribution from the $q$ is obtained by multiplying a symmetric factor at this order.

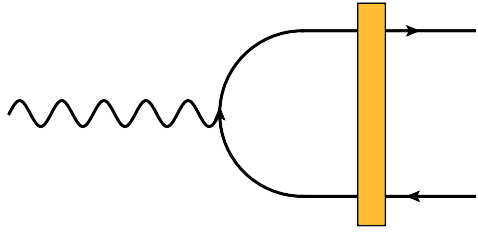

FIG. 1. Dipole amplitude for $\gamma^* \to q\bar{q}$. The orange block represents the CGC Wilson line.

We work in the Breit frame, in which $q = (0, 0, 0, -Q)$ for the virtual photon, and $P = \frac{Q}{2x_B}(1, 0, 0, 1)$ for the nucleus. The calculation is straightforward which gives

$$\Sigma(Q^2, x_B, \theta) = \sum_{\lambda=l,t} f_\lambda \Sigma_\lambda^{\gamma^*}(Q^2, x_B, \theta), \tag{2}$$

where we sum over the virtual photon polarization with the flux $f_t = 1 - y + \frac{y^2}{2}$ for the transverse and $f_l = 1 - y$ longitudinal polarization, respectively, with $y = \frac{Q^2}{x_B s}$ the inelasticity. Here for the transverse photon contribution, we find

$$\Sigma_t^{\gamma^*}(Q^2, x_B, \theta) = \sum_q \frac{2N_c \alpha^2 e_q^2}{\pi^2 x_B Q^2} S_\perp \int dz d^2\vec{k}_t \frac{d^2\vec{l}_t}{(2\pi)^2} F_{g,x_B}(\vec{l}_t) \left[z^2 + (1-z)^2\right] \left| \frac{\vec{k}_t}{\vec{k}_t^2 + \Delta^2} - \frac{\vec{k}_t - \vec{l}_t}{(\vec{k}_t - \vec{l}_t)^2 + \Delta^2} \right|^2$$
$$\times \left[ \frac{\vec{k}_t^2 + (1-z)^2 Q^2}{(1-z)Q} \frac{x_B}{Q} \right] \frac{1}{2\theta} \delta\left(\theta - \tan^{-1}\frac{2k_t(1-z)Q}{k_t^2 - (1-z)^2 Q^2}\right) \theta\left(\frac{\vec{k}_t^2 + (1-z)^2 Q^2}{(1-z)Q} < x_B \frac{Q}{x_B}\right), \tag{3}$$

where $k$ is the momentum for $\bar{q}$ and $1 - z = \frac{k^-}{Q}$ the momentum fraction with respect to the photon. We defined $\Delta^2 = z(1-z)Q^2$. In the last line, the first term is the energy weight, derived using the on-shell condition $k^+ k^- - \vec{k}_t^2 = 0$,

---

* xiliu@bnu.edu.cn
† fyuan@lbl.gov
‡ zhuhx@zju.edu.cn

normalized to the incoming nucleus energy $Q/x_B$ in the Breit frame. The second term defines the $\theta$ angle with respect to the $z$-axis, with $\theta < \pi/2$ for $k_z > 0$ while $\theta > \pi/2$ for $k_z < 0$. The last $\theta$ function ensures that $2k^0$ can not exceed the incoming parton momentum $x_B P = x_B \frac{Q}{x_B}$ [1]. As for the longitudinal polarization, we have

$$\Sigma_l^{\gamma^*}(Q^2, x_B, \theta) = \sum_q \frac{2N_c \alpha^2 e_q^2}{\pi^2 x_B Q^2} S_\perp \int dz d^2\vec{k}_t \frac{d^2\vec{l}_t}{(2\pi)^2} F_{g,x_B}(\vec{l}_t) 4z^2(1-z)^2 Q^2 \left| \frac{1}{\vec{k}_t^2 + \Delta^2} - \frac{1}{(\vec{k}_t - \vec{l}_t)^2 + \Delta^2} \right|^2$$
$$\times \left[ \frac{\vec{k}_t^2 + (1-z)^2 Q^2}{(1-z)Q} \frac{x_B}{Q} \right] \frac{1}{2\theta} \delta\left( \theta - \tan^{-1} \frac{2k_t(1-z)Q}{k_t^2 - (1-z)^2 Q^2} \right) \theta\left( \frac{\vec{k}_t^2 + (1-z)^2 Q^2}{(1-z)Q} < x_B \frac{Q}{x_B} \right). \quad (4)$$

Now we consider the small-$\theta$ limit. As mentioned above, is the momentum "+"-component that enters the detector is $k^+ = \frac{(1-\xi)}{\xi} Q$, meanwhile it can also be written as $k^+ = \frac{k_t^2}{Q(1-z)}$ by the on-shell condition. Thus, the variable $\xi$ can be related to $Q$ and $z$ through

$$(1-z)Q = \frac{\xi}{1-\xi} \frac{k_t^2}{Q}. \quad (5)$$

Moreover, in the collinear limit, the magnitude of the transverse momentum $k_t$ can be defined in terms of the "+"-component and its polar angle $\theta_k$ by

$$k_t = \frac{Q}{2} \frac{1-\xi}{\xi} \theta_k, \quad (6)$$

and thus

$$(1-z)Q = \frac{1}{Q} \frac{Q^2}{4} \frac{1-\xi}{\xi} \theta_k^2. \quad (7)$$

The integral measure $dz d^2\vec{k}_t$ can then be written as

$$dz d^2\vec{k}_t = \frac{k_t^2 d\xi}{Q^2(1-\xi)^2} \frac{k_t^2}{\theta_k} d\theta_k d\phi. \quad (8)$$

We note that here $k_t$ is given by Eq. (6).

Furthermore, the relations in Eq. (6) and Eq. (7) suggest that when $\theta_k \to 0$ is small, the transverse momentum $k_t$ is approaching to 0 while $z \to 1$. Therefore in the small angle limit, we will only keep the leading contribution to the cross section in Eq. (3) and Eq. (4) in the $z \to 1$ limit. It is ready to see that $\Sigma_l^{\gamma^*} \to 0$ in this limit, while $\Sigma_t$ reduces to

$$\Sigma_t^{\gamma^*}(Q^2, x_B, \theta) = \sum_q \frac{2N_c \alpha^2 e_q^2}{\pi^2 x_B Q^2} S_\perp \int \frac{k_t^2 d\xi}{Q^2(1-\xi)^2} \frac{k_t^2}{\theta_k} d\theta_k d\phi \frac{d^2\vec{l}_t}{(2\pi)^2} F_{g,x_B}(\vec{l}_t) [1] \left| \frac{\vec{k}_t}{\vec{k}_t^2 + \frac{\xi}{1-\xi}\vec{k}_t^2} - \frac{\vec{k}_t - \vec{l}_t}{(\vec{k}_t - \vec{l}_t)^2 + \frac{\xi}{1-\xi}\vec{k}_t^2} \right|^2$$
$$\times \left[ Q \frac{1-\xi}{\xi} \frac{x_B}{Q} \right] \frac{1}{2\theta} \delta(\theta - \theta_k) \theta\left( Q \frac{1-\xi}{\xi} < x_B \frac{Q}{x_B} \right), \quad (9)$$

where we have applied Eq. (8) to replace the phase space measure in Eq. (3) and used the fact that when $z \to 1$

$$(1-\xi)\Delta^2 = (1-\xi)Q^2 z(1-z) \approx (1-\xi) \frac{\xi}{1-\xi} k_t^2 = \xi k_t^2, \quad (10)$$

to substitute the $\Delta^2$ in the first line of Eq. (3). Meanwhile the second line of Eq. (3) is simplified by Eq. (6) such that $\frac{\vec{k}_t^2}{(1-z)Q} = \frac{1-\xi}{\xi} Q$ and ignoring all terms proportional to $(1-z)Q$. For instance for $\frac{\vec{k}_t^2 + (1-z)^2 Q^2}{(1-z)Q}$, we have

$$\frac{\vec{k}_t^2 + (1-z)^2 Q^2}{(1-z)Q} \approx \frac{\vec{k}_t^2}{(1-z)Q} = \frac{1-\xi}{\xi} Q. \quad (11)$$

Perform the $\theta_k$ integration using the $\delta(\theta - \theta_k)$ we arrive at the final result for $\Sigma(Q^2, x_B, \theta)$ when $\theta \to 0$

$$\Sigma_t^{\gamma^*}(Q^2, x_B, \theta) = \sum_q \frac{2N_c \alpha^2 e_q^2}{\pi^2 Q^2} S_\perp \int \frac{d\xi}{Q^2} \frac{1}{2\theta^2} \frac{d^2 \vec{l}_t}{2\pi} k_t^2 k_t^2 \left| \frac{\vec{k}_t}{k_t^2} - \frac{\vec{k}_t - \vec{l}_t}{(1-\xi)(\vec{k}_t - \vec{l}_t)^2 + \xi k_t^2} \right|^2$$

$$\times \left[ \frac{1-\xi}{\xi} \right] \theta\left( \frac{1-\xi}{\xi} < 1 \right) F_{g, x_B}(\vec{l}_t),$$

$$= \sum_q \frac{4\pi \alpha^2 e_q^2}{Q^4} \frac{N_c S_\perp}{8\pi^4} \frac{1}{\theta^2} \int \frac{d\xi}{\xi} d^2 \vec{l}_t (1-\xi) k_t^2 (\vec{k}_t - \vec{l}_t)^2 \left| \frac{\vec{k}_t}{\xi \vec{k}_t^2 + (1-\xi)(\vec{k}_t - \vec{l}_t)^2} - \frac{\vec{k}_t - \vec{l}_t}{(\vec{k}_t - \vec{l}_t)^2} \right|^2$$

$$\times \theta\left( \frac{1-\xi}{\xi} < 1 \right) F_{g, x_B}(\vec{l}_t), \tag{12}$$

where we have applied

$$k_t^2 \left| \frac{\vec{k}_t}{k_t^2} - \frac{\vec{k}_t - \vec{l}_t}{(1-\xi)(\vec{k}_t - \vec{l}_t)^2 + \xi k_t^2} \right|^2$$

$$= 1 + \frac{k_t^2 (\vec{k}_t - \vec{l}_t)^2}{((1-\xi)(\vec{k}_t - \vec{l}_t)^2 + \xi k_t^2)^2} - 2 \frac{\vec{k}_t \cdot (\vec{k}_t - \vec{l}_t)}{((1-\xi)(\vec{k}_t - \vec{l}_t)^2 + \xi k_t^2)}$$

$$= (\vec{k}_t - \vec{l}_t)^2 \left[ \frac{(\vec{k}_t - \vec{l}_t)^2}{(\vec{k}_t - \vec{l}_t)^4} + \frac{k_t^2}{((1-\xi)(\vec{k}_t - \vec{l}_t)^2 + \xi k_t^2)^2} - 2 \frac{\vec{k}_t \cdot (\vec{k}_t - \vec{l}_t)}{(\vec{k}_t - \vec{l}_t)^2((1-\xi)(\vec{k}_t - \vec{l}_t)^2 + \xi k_t^2)} \right]$$

$$= (\vec{k}_t - \vec{l}_t)^2 \left| \frac{\vec{k}_t}{\xi \vec{k}_t^2 + (1-\xi)(\vec{k}_t - \vec{l}_t)^2} - \frac{\vec{k}_t - \vec{l}_t}{(\vec{k}_t - \vec{l}_t)^2} \right|^2. \tag{13}$$

We thus identify $f_{\text{EEC}}(x_B, \theta)$ at this order in the small-$x$ formalism as

$$f_{q,\text{EEC}}(x_B, \theta) = \frac{N_c S_\perp}{8\pi^4} \frac{1}{\theta^2} \int_{\xi_{\text{cut}}}^1 \frac{d\xi}{\xi} d^2 \vec{l}_t (1-\xi) k_t^2 (\vec{k}_t - \vec{l}_t)^2 \left| \frac{\vec{k}_t}{\xi \vec{k}_t^2 + (1-\xi)(\vec{k}_t - \vec{l}_t)^2} - \frac{\vec{k}_t - \vec{l}_t}{(\vec{k}_t - \vec{l}_t)^2} \right|^2 F_{g, x_B}(\vec{l}_t). \tag{14}$$

The $\xi_{\text{cut}}$ is determined from the last $\theta$-function in Eq. (9).

---



\end{document}